\documentclass[aps,twocolumn]{revtex4}
\usepackage{graphicx}

\begin{document}

\title{Decoupling of itinerant and localized $d$-orbital electrons in the compound Sc$_{0.5}$Zr$_{0.5}$Co}
\author{Jin Si$^{1}$, Xinwei Fan$^{1}$, Enyu Wang$^{2}$, Xiyu Zhu$^{1 *}$, Qing Li$^{1}$, Hai-Hu Wen$^{1}$}
\email{zhuxiyu@nju.edu.cn,hhwen@nju.edu.cn}

\affiliation{$^{1}$Center for Superconducting Physics and Materials, National Laboratory of Solid State Microstructures and Department of Physics, National Center of Microstructures and Quantum Manipulation, Nanjing University, Nanjing 210093, China}
\affiliation{$^{2}$School of Physics and Electronic Engineering, Linyi University, Linyi 276000, China}

\begin{abstract}
By using the arc-melting method, we successfully synthesized the compound Sc$_{0.5}$Zr$_{0.5}$Co with the space group of $Pm$-$3m$. Both the resistivity and magnetic susceptibility measurements reveal a phase transition at about 86 K. This transition might be attributed to the establishment of an antiferromagnetic order. The magnetization hysteresis loop measurements in wide temperature region show a weak ferromagnetic feature, which suggests a possible canted arrangement of the magnetic moments. Bounded by the phase transition temperature, the resistivity at ambient pressure shows a change from Fermi liquid behavior to a super-linear behavior as temperature increases. By applying pressures up to 32.1 GPa, the transition temperature does not show a clear change and no superconductivity is observed above 2 K. The density functional theory (DFT) calculations confirm the existence of the antiferromagnetic order and reveal a gap between the spin-up and spin-down $d$-orbital electrons. This kind of behavior may suggest that the antiferromagnetic order in this compound originates from the localized $d$-electrons which do not contribute to the conductance. Thus the itinerant and localized $d$-orbital electrons in the compound are decoupled.
\end{abstract}

\maketitle

\section{Introduction}

Since the discovery of cuprates\cite{Cuprate} and iron based superconductors\cite{Iron}, antiferromagnetic spin fluctuations are assumed to be the possible superconducting pairing glue, which leads to the unconventional superconductivity\cite{glue}. And the emergence of superconductivity accompanied by the suppression of long-range antiferromagnetic order becomes one of the common features of unconventional superconductors, such as cuprates, iron based superconductors, and CrAs, etc.\cite{Cuprate,IronAFM,CrAs}. It has been known that, for CrAs and some iron based parent compounds like LaFeAsO and BaFe$_{2}$As$_{2}$, high pressure is an effective and clean way to suppress the antiferromagnetic order and induce superconductivity\cite{parent1,parent2}. However, high pressure is not always effective in bring in superconductivity for some compounds with the antiferromagnetic order originated from localized $d$-electrons, such as NiO\cite{NiO}, and Cr$_{2}$GaN\cite{Cr2GaN}. For this kind of compounds, it is very hard to suppress the antiferromagnetic order solely by external pressures.

Nowadays, alloys containing cobalt have attracted much attention, because these compounds sometimes show novel physical properties which are related to the correlation effect generated from 3$d$-electrons of cobalt. For example, the cobalt based Heusler alloy Co$_2$MnSi exhibits a huge spin polarization, and Co$_2$CrAl reveals a temperature-dependent anomalous Hall effect\cite{Co2MnSi1,Co2MnSi2,Co2CrAl}. The compound Zr$_{2}$Co was reported as a superconductor with an itinerant antiferromagnetism in the normal state revealed by the NMR study\cite{Zr2Co,NMR}. Meanwhile, ScZrCo alloy crystallized in the Ti$_{2}$Ni structure, was reported as a correlated bad metal and could become a superconductor by the high pressure tuning\cite{ScZrCo}.

Here, we report a compound Sc$_{0.5}$Zr$_{0.5}$Co with a possible antiferromagnetic transition at about 86 K. And the phase transition seems to be quite robust under high pressures. Only in the low pressure region, for example below 3.1 GPa, there is a slight suppression of the antiferromagnetic transition temperature from 86 K at ambient pressure to 74.3 K under 3.1 GPa. However, with further increasing the pressure, the antiferromagnetic transition temperature becomes unchangeable and no superconductivity is observed above 2 K. This feature probably indicates the localized $d$-electrons, which are not involved in conducting, are the reason for the antiferromagnetic order in Sc$_{0.5}$Zr$_{0.5}$Co. The DFT calculations of this compound indicate it indeed has an antiferromagnetic ground state and there is a huge gap between the band of the spin-up and spin-down $d$-orbital electrons originated from the Co atoms along the X-W line, which means spin-flip can hardly happen with the external magnetic field.

\section{Experimental details}

By using the arc-melting method, we fabricated the polycrystalline samples of Sc$_{0.5}$Zr$_{0.5}$Co. Firstly, the scandium(Aladdin, ingot, purity 99.9\%) and zirconium (Aladdin, slug, purity 99.5\%) were crashed and grounded into powders. Then, the scandium, zirconium and cobalt powders (Alfa Aesar, powder, purity 99.99\%) with the mole ratio of 1:1:2 were weighed, ground carefully and pressed into a pellet in a glove box filled with argon. Afterwards, the pellet was melted in a copper base arc-melting furnace. During the arc-melting process, the chamber was evacuated first and then it was filled with Ar gas. To improve the homogeneity of samples, each pellet was re-melted at least four times.

The X-ray-diffraction (XRD) measurements were carried out on a Bruker D8 Advanced diffractometer with the CuK${_{\alpha1}}$  radiation. The energy dispersive spectrums were obtained with a Phenom ProX scanning electron microscope(SEM) at an accelerating voltage of 15 kV. The resistivity at ambient pressure was measured with the standard four-probe method on a Physical Property Measurement System(PPMS, Quantum Design). The DC magnetization measurements were performed on a SQUID-VSM (Quantum Design). The resistivity data under high pressure was collected by using a diamond-anvil-cell (DAC) module which matches our PPMS (cryoDAC-PPMS, Almax easyLab) equipment with a four-probe van der Pauw method\cite{VanDerPauw}. The applied pressure was determined by measuring the shift of ruby R$_1$ luminescent line\cite{Ruby}.

The density functional theory (DFT) calculations were performed by the WIEN2k package\cite{Wien2k}, which utilizes the full-potential linearized augmented plane-wave in contrast to the pseudo-potential method. The Perdew-Burke-Ernzerhof (PBE) exchange correlation energy was implemented to carry out the self-consistent field cycle. The convergence criterions were $10^{-4}$ Ry for energy and $10^{-3}$ e for charge.

\section{Results and discussion}
\subsection{Sample characterization}

\begin{figure}
\centering\includegraphics[width=8.5cm,height=9.5cm]{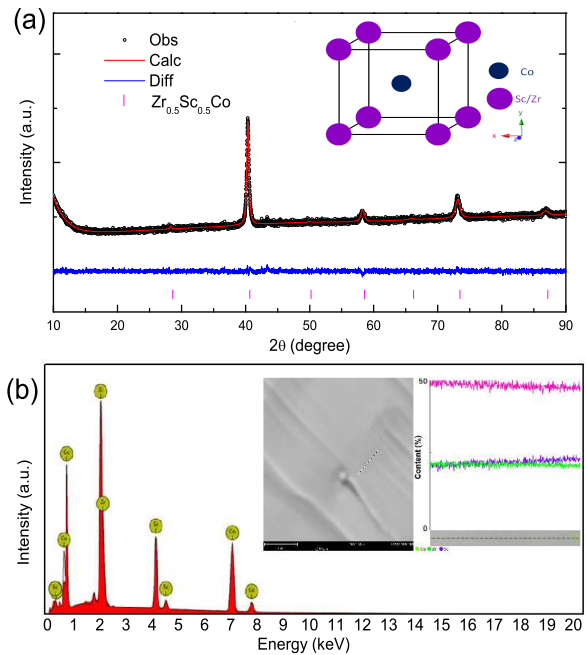}
\caption{(Color online)(a) X-ray diffraction pattern of the Sc$_{0.5}$Zr$_{0.5}$Co polycrystalline sample measured at room temperature. The inset shows the schematic structure of Sc$_{0.5}$Zr$_{0.5}$Co. (b) EDS line scan on SEM image and the content ratio of Co, Zr and Sc.}\label{fig1}
\end{figure}

Fig.\ref{fig1}(a) presents the XRD pattern (symbols) for the Sc$_{0.5}$Zr$_{0.5}$Co polycrystalline sample with the Rietveld refinement\cite{Rietveld} of the data shown by the red curve. The Rietveld refinements are conducted with the TOPAS 4.2 software\cite{TOPAS}. The Rietveld refinement agreement factors R$_{wp}$ = 2.47\% and R$_{p}$ = 1.95\% are obtained and found relatively small, which means that the calculated profile agrees with our experimental data quite well. The lattice parameters are calculated to be $a$ = $b$ = $c$ = 3.178(5) \r{A} with the space group of $Pm$-$3m$. More details of the Rietveld refinement profiles are shown in Table~\ref{tbl}. The inset of Fig.\ref{fig1}(a) shows the schematic structure of Sc$_{0.5}$Zr$_{0.5}$Co, in which the Co atoms occupy one position; Sc and Zr atoms occupy other positions probably in a random way. In order to determine the exact element ratio of the title compound, we also measure a line scan of Energy Dispersive Spectroscopy (EDS), which is in Fig.\ref{fig1}(b). As we can see, the ratio of the three elements is very uniform with Sc : Zr : Co = 1 : 0.97 : 2.08, which is close to the nominal compositions.

\begin{table}
  \centering
  \caption{Crystallographic data of Sc$_{0.5}$Zr$_{0.5}$Co at 300 K.}
  \label{tbl}
  \begin{tabular}{cccccccc}
    \hline
    \multicolumn{3}{l}{compound}  & \multicolumn{3}{l}{Sc$_{0.5}$Zr$_{0.5}$Co}  \\
    \hline
    \multicolumn{3}{l}{space group}          & \multicolumn{3}{l}{$Pm-3m$}   \\
    \multicolumn{3}{l}{$a$ ($\mathring{A}$) }    & \multicolumn{3}{l}{3.178(5)}  \\
    \multicolumn{3}{l}{$V$ ($\mathring{A}^3$)  }   & \multicolumn{3}{l}{32.112(1)}  \\
    \multicolumn{3}{l}{$\rho$ (g/cm$^3$)  }   & \multicolumn{3}{l}{6.568}  \\
    \multicolumn{3}{l}{$R_{wp}$ (\%) }           & \multicolumn{3}{l}{2.47} \\
    \multicolumn{3}{l}{$R_{p}$ (\%) }           & \multicolumn{3}{l}{1.95} \\
    \multicolumn{3}{l}{$GOF$}           & \multicolumn{3}{l}{1.06} \\
    \hline
    atom & site & x & y & z & occupancy & B$_{eq}$ \\
    \hline
    Co  & 1b & 0.5 & 0.5 & 0.5 & 1 & 1.464 \\
    Sc & 1a & 0 & 0 & 0 & 0.5 & 0.1325 \\
    Zr & 1a & 0 & 0 & 0 & 0.5 & 0.8759 \\
    \hline

  \end{tabular}
\end{table}

\subsection{Transport and magnetic properties}

\begin{figure}
\centering\includegraphics[width=8cm,height=10.6cm]{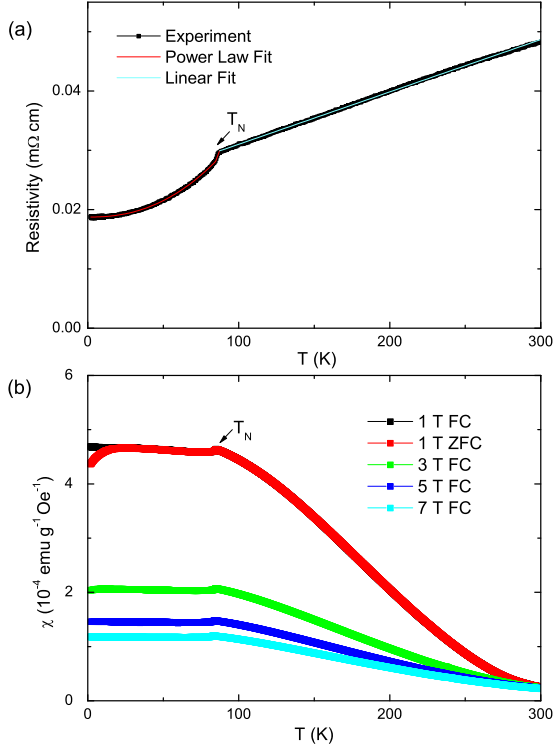}
\caption{(Color online) Temperature dependence of (a) resistivity and (b) magnetic susceptibility with different magnetic fields for Sc$_{0.5}$Zr$_{0.5}$Co at ambient pressure.} \label{fig2}
\end{figure}

Fig.\ref{fig2}(a) shows the temperature dependence of resistivity for Sc$_{0.5}$Zr$_{0.5}$Co at ambient pressure. As we can see, a clear anomalous kink occurs at about 86 K indicating a possible phase transition. Above this temperature, the resistivity shows a linear behavior with the resistance temperature coefficient of about 0.0878 $\mu$$\Omega$$\cdot$cm/K. However, below the phase transition temperature, the resistivity curve reveals a quadratic relation, which is the typical feature of a Fermi liquid behavior. In Fig.\ref{fig2}(b), we present the zero-field cooled (ZFC) and field cooled (FC) DC magnetization for Sc$_{0.5}$Zr$_{0.5}$Co sample under several magnetic fields. A kink at the same temperature shown in the resistivity is also detected. Around 86 K, the magnetic susceptibility data reveals a clear signature of phase transition. The maximum of magnetic susceptibility with external field of 1 T in the SI system is about 0.03, which is much smaller than that in normal ferromagnetic materials (10-10$^6$). This kind of phase transition behavior is very similar to some antiferromagnetic compounds like TiAu, BaFe$_{2}$As$_{2}$, et al.\cite{TiAu,BaFe2As2}. Thus, the phase transition at 86 K probably corresponds to an antiferromagnetic transition. And if it is an AF ordering temperature, the N\'eel temperature does not change when we increase the magnetic fields up to 7 T. Besides these, in the low temperature region below 15 K at 1 T, $\chi$$_{dc}$(T) shows a deviation between the ZFC and FC curves, which indicates a hysteresis of magnetization and may suggest a spin-glass-like behavior.

\begin{figure}
\centering\includegraphics[width=8cm,height=6.5cm]{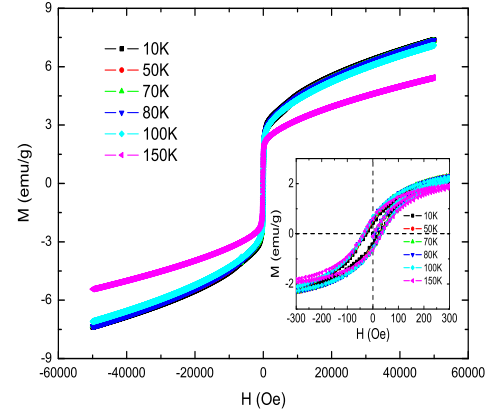}
\caption{(Color online) The magnetization hysteresis loop (MHL) measurements for the Sc$_{0.5}$Zr$_{0.5}$Co sample at different temperatures. The inset shows the details of MHL measurements at low magnetic field.} \label{fig3}
\end{figure}

The magnetization hysteresis loop (MHL) measurements at different temperatures shown in Fig.\ref{fig3} exhibit a weak ferromagnetic feature. And the feature does not change disregard the temperature is above or below the N\'eel temperature. It may suggest that the magnetic moments of Sc$_{0.5}$Zr$_{0.5}$Co have a canted feature leading to the weak ferromagnetism. The coercive magnetic field H$_C$ and remanent magnetization M$_r$ are detailed in the inset of Fig.\ref{fig3}. Because the antiferromagnetic order is mainly caused by the neighboring Co atoms with opposite spins, some vacancies of Co atoms or the unavoidable mutual doping of three kinds of atoms are thought to be responsible for the weak ferromagnetic feature. However, we cannot rule out the possibility that the weak ferromagnetic moment comes from the effect of magnetic impurities, although we cannot see any trace of a second phase from the XRD pattern.

\begin{figure}
\centering\includegraphics[width=10cm,height=8cm]{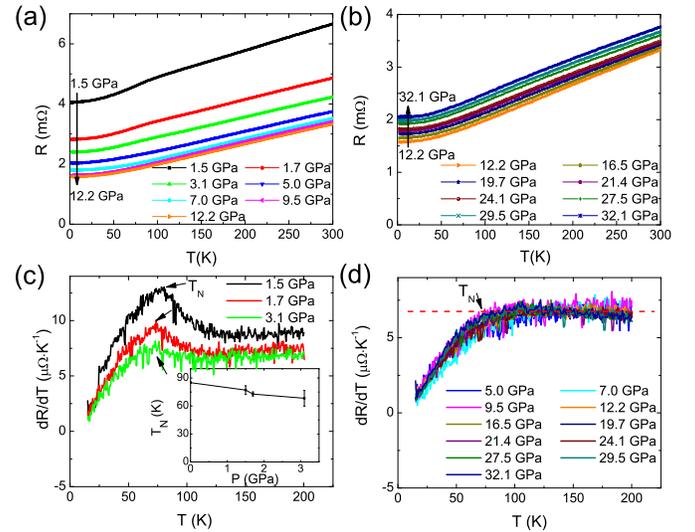}
\caption{(Color online) Temperature dependence of resistance under different pressures: (a) 1.5 to 12.2 GPa, (b) 12.2 to 32.1 GPa and derivative dR/dT under pressures: (c) 1.5 to 3.1 GPa, (d) 3.1 to 32.1 GPa for the Sc$_{0.5}$Zr$_{0.5}$Co sample in the temperature range from 2 K to 300 K. The inset in (c) shows the pressure dependence of T$_N$ from 1.5 to 3.1 GPa} \label{fig4}
\end{figure}

Fig.\ref{fig4}(a) and (b) display the temperature dependence of resistance for Sc$_{0.5}$Zr$_{0.5}$Co under different external pressures, and no superconductivity is observed up to 32.1 GPa above 2 K. In the beginning, as increasing the pressures, the resistance of the sample decreases until 12.2 GPa, together with the enlarged Residual Resistance Ratio (${RRR}$). With further increasing pressure, the resistance gets enhanced slowly, while the ${RRR}$ slightly decreases. In order to determine the phase transition temperature more accurately, we take the derivative of resistance with respect to temperature and show the results in Fig.\ref{fig4}(c) and (d). The anomalies of the dR/dT curves as marked here represent the antiferromagnetic transition. We take the temperatures of kink as the N\'eel temperatures. Below 3.1 GPa, the N\'eel temperature slowly decreases with increasing pressure, as shown in the inset of Fig.\ref{fig4}(c), it changes from 86 K at ambient pressure to 74.3 K at 3.1 GPa. However, when we keep increasing the pressure, as shown in Fig.\ref{fig4}(d), the N\'eel temperature does not change and remains at 74.3 K. These indicate that the antiferromagnetic order is very robust and the response to pressure is very weak in Sc$_{0.5}$Zr$_{0.5}$Co. However, as we can conclude from Fig.\ref{fig4}(a) and (b), the conductance which is attributed by itinerant electrons can be easily tuned by pressure. Thus, the electrons attributing to the antiferromagnetic order should be localized. In other words, the itinerant $d$-orbital electrons contributing to the conductance are decoupled with the localized $d$-electrons which is responsible for the antiferromagnetic order. Therefore, this may be the reason why we cannot suppress the antiferromagnetic order and obtain possible superconductivity in this compound.

\begin{figure}
\centering\includegraphics[width=8cm,height=6.5cm]{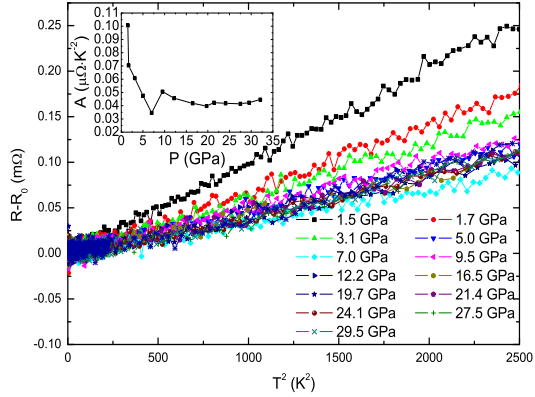}
\caption{(Color online) The difference of resistance versus the square of temperature under different pressures in the temperature region from 2 K to 50 K. The inset shows the pressure dependence of the fitting parameter $A$ using the formula $R$ = $R_0$+$A T^2$. } \label{fig5}
\end{figure}

Fig.\ref{fig5} shows the fitting results to the resistance curves by using the formula $R$ = $R_0$+$A T^2$ under different pressures. As we can see, under all pressures, the resistance can be well fitted by the formula with a quadratic temperature dependence, which indicates that the system shows a a persistent Fermi liquid behavior at all pressures. As shown in the inset of Fig.\ref{fig5}, the fitting parameter $A$ shows a linear change from 1.7 GPa to 7.0 GPa with a rate of -6.754 n$\Omega$/(K$^2$$\cdot$GPa). While, above 12.2 GPa, $A$ remains basically unchanged. This is consistent with the response of the conductance to the pressure. Below 12.2 GPa, conductivity enhances with increasing pressure. Further increasing pressure, the conductivity is weakly affected by pressure.

\subsection{Ab-initio Calculations and Discussions}

\begin{figure}
\centering\includegraphics[width=9cm,height=13.5cm]{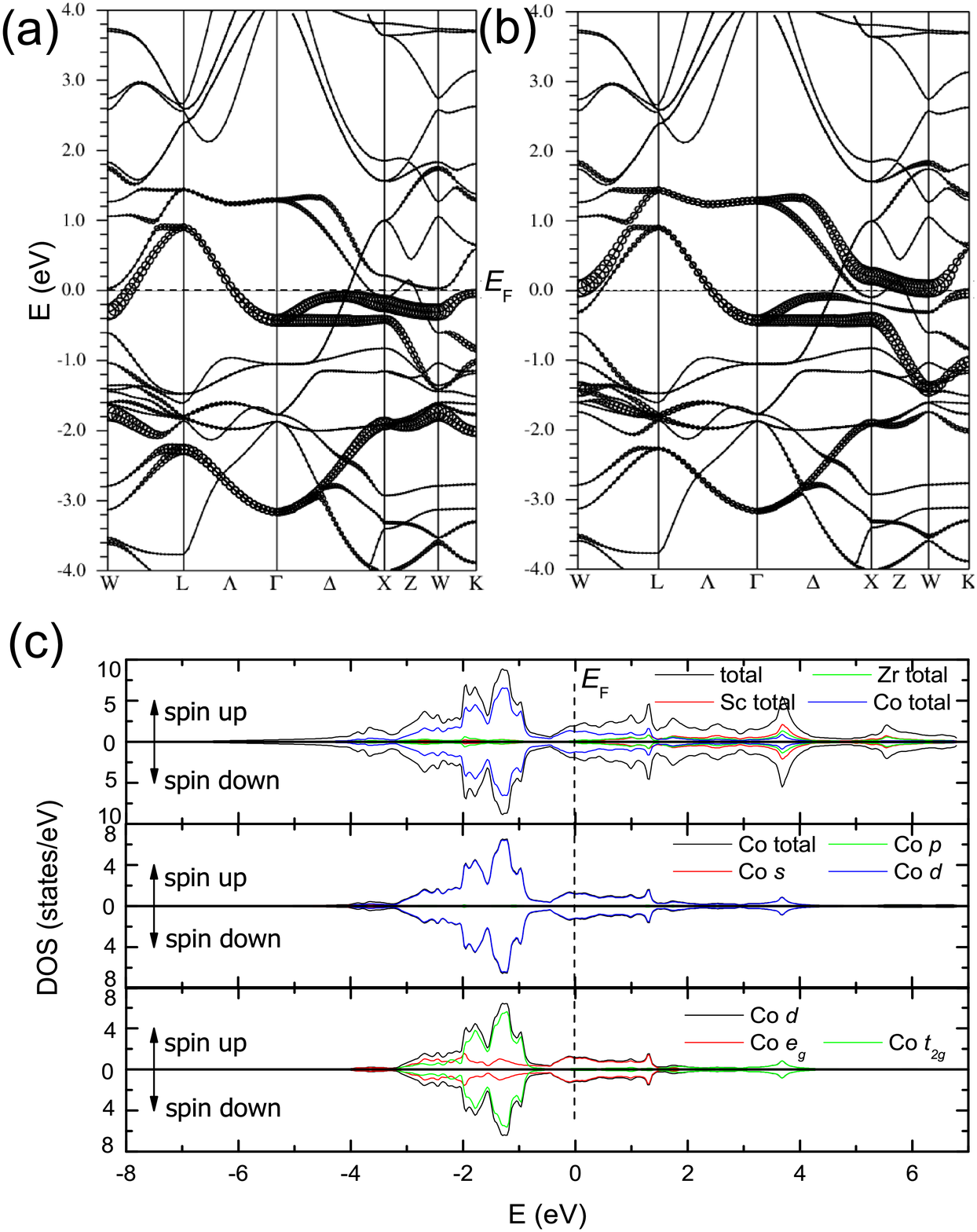}
\caption{(Color online) Band Structure of Sc$_{0.5}$Zr$_{0.5}$Co for (a)spin up and (b)spin down. The size of the symbols represents the weight of $e$$_g$-orbital electrons of Co1 atoms. (c)Total and orbital projected DOS of Sc$_{0.5}$Zr$_{0.5}$Co. } \label{fig6}
\end{figure}

In order to have a comprehensive understanding to the data obtained in this experiment, we have conducted ab initio DFT calculations. We think the antiferromagnetic order probably comes from the Co atoms and divide the neighboring Co atoms into two parts: Co1 atoms and Co2 atoms and their magnetic moment are arranged antiferromagnetically. From the band structure and the density of states (DOS) diagram shown in Fig.\ref{fig6}, we can see that the DOS for both spins are almost identical, which suggests that the ferromagnetic order should not be the ground state for this compound. Also, the spin magnetic moment is 0.304 $\mu$$_B$ per Co1 atom and -0.304 $\mu$$_B$ per Co2 atom, which indicates an antiferromagnetic ground state. A closer scrutinizing to band structures along the X-W line, the fat bands near the Fermi energy consist of localized $e$$_g$-orbital electrons of Co atoms. And there is a clear gap about 0.34 eV between the up spins and down spins, which should be responsible for the robust antiferromagnetic properties we observed in experiments. Meanwhile, some itinerant $e$$_g$-orbital electrons of Co atoms form a hole pocket at L point, and they are the main part of the itinerant electrons contributing to the conductivity. Therefore it is clear that the electrons from different sub-orbits play different roles. Some are localized and responsible for the formation of magnetic moment, while the other are itinerant leading to the electrical conduction. In the transition metal compounds, the balance between the localization and itineracy of the $d$-electrons is very essential to determine the intrinsic properties of the materials.

\section{Conclusions}
In conclusion, we have successfully synthesized Sc$_{0.5}$Zr$_{0.5}$Co by arc-melting method. The magnetization and resistivity data indicate a possible antiferromagnetic transition at 86 K, although a weak ferromagnetic signal is detected in the temperature regions both below and above the AF transition temperature. By applying external pressures, the conductance of the sample is significantly affected, but the antiferromagnetic transition temperature responses weakly. Our DFT calculations indicate that the $d$ electrons of Co atoms split into itinerant part and localized part. The itinerant part is responsible for the conductivity of the compound, while the localized part gives rise to the antiferromagnetism we observed in experiments. Besides, we can see from the band structures that there is a clear gap between the up and down spins, which indicates a strong antiferromagnetism and that is consistent with our magnetization measurements.

\section*{ACKNOWLEDGMENTS}
This work was supported by the National Key R\&D Program of China (Grant nos. 2016YFA0300401 and 2016YFA0401704) and National Natural Science Foundation of China (Grant nos. A0402/11534005 and A0402/11674164).


\begin{thebibliography}{00}

\bibitem{Cuprate}J. G. Bednorz and K. A. M\"uller, Z. Phys. B. \textbf{64}, 189-193 (1986).
\bibitem{Iron}Y. Kamihara, H. Hiramatsu, M. Hirano, R. Kawamura, H. Yanagi, T. Kamiya and H. Hosono, J. Am. Chem. Soc. \textbf{128}, 10012-10013 (2006).
\bibitem{glue}N. D. Mathur, F. M. Grosche, S. R. Julian, I. R. Walker, D. M. Freye, R. K. W. Haselwimmer and G. G. Lonzarich,  Nature \textbf{394}, 39-43 (1998).
\bibitem{IronAFM}C. de la Cruz, Q. Huang, J. W. Lynn, J. Li, W. Ratcliff II, J. L. Zarestky, H. A. Mook, G. F. Chen, J. L. Luo, N. L. Wang and Pengcheng Dai, Nature (London) \textbf{453}, 899 (2008).
\bibitem{CrAs}W. Wu, J. Cheng, K. Matsubayashi, P. Kong, F. Lin, C. Jin, N. Wang, Y. Uwatoko and J. Luo, Nature Communications \textbf{5}, 5508 (2014).
\bibitem{parent1}H. Okada, K. Igawa, H. Takahashi, Y. Kamihara, M. Hirano, H. Hosono, K. Matsubayashi and Y. Uwatokod, J. Phys. Soc. Jpn. \textbf{77}, 113712 (2008).
\bibitem{parent2}P. Alireza, Y. Ko, J. Gillett, C. Petrone, J. Cole, G. Lonzarich and S. Sebastian, J. Phys.: Condens. Matter \textbf{21}, 012208 (2009).
\bibitem{NiO}V. Potapkin, L. Dubrovinsky, I. Sergueev, M. Ekholm, I. Kantor, D. Bessas, E. Bykova, V. Prakapenka, R. P. Hermann, R. Ruffer, V. Cerantola, H. J. M. Jonsson, W. Olovsson, S. Mankovsky, H. Ebert and I. A. Abrikosov Phys. Rev. B \textbf{93}, 201110(R) (2016).
\bibitem{Cr2GaN}Y. Li, J. Liu, W. Liu, X. Zhu and H. H. Wen, Philosophical Magazine \textbf{95}, 2773-2780 (2015).
\bibitem{Co2MnSi1}Y. Sakuraba, J. Nakata, M. Oogane, H. Kubota, Y. Ando, A. Sakuma and T. Miyazaki, Jpn. J. Appl. Phys. \textbf{44}, L1100 (2005).
\bibitem{Co2MnSi2}M. Jourdan, J. Min\'ar, J. Braun, A. Kronenberg, S. Chadov, B. Balke, A. Gloskovskii, M. Kolbe,
H.J. Elmers, G. Sch\"onhense, H. Ebert, C. Felser and M. Kl\"aui, Nature Communications \textbf{5}, 3974 (2014).
\bibitem{Co2CrAl}A. Husmann and L. J. Singh, Phys. Rev. B \textbf{73}, 172417 (2006).
\bibitem{Zr2Co}S. L. McCarthy, J. Low Temp. Phys. \textbf{4}, 489-501 (1971).
\bibitem{NMR}R. Miller, and C. Satterthwaite, Phys. Rev. B \textbf{58}, 11698-11702 (1998).
\bibitem{ScZrCo}E. Wang, J. Si, X. Zhu, G. Chen and H. H. Wen, New J. Phys. \textbf{20}, 073036 (2018).
\bibitem{VanDerPauw}L.J. van der Pauw, Philips Res.Rep. \textbf{13}, 1-9 (1958).
\bibitem{Ruby}H. K. Mao, J. Xu, and P.M. Bell, J. Geophys. Res. Solid Earth \textbf{91}, 4573-4676 (1986).
\bibitem{Wien2k}P. Blaha, K. Schwarz, G. K. H. Madsen, D. Kvasnicka, J. Luitz, R. Laskowski, F. Tran and L. D. Marks, WIEN2k, An Augmented Plane Wave + Local Orbitals Program for Calculating Crystal Properties (Karlheinz Schwarz, Techn. Universit\"at Wien, Austria), ISBN 3-9501031-1-2 (2018).
\bibitem{Rietveld}H. M. Rietveld, J. Appl. Crystallogr. \textbf{2}, 65-71 (1969)
\bibitem{TOPAS}R. W. Cheary and A. Coelho, J. Appl. Crystallogr. \textbf{25}, 109-121 (1992)
\bibitem{TiAu}E. Svanidze, J. K. Wang, T. Besara, L. Liu, Q. Huang, T. Siegrist, B. Frandsen, J. W. Lynn, A. H. Nevidomskyy, M. B. Gamza, M. C. Aronson, Y. J. Uemura and E. Morosan, Nature Communications \textbf{6}, 7701 (2015).
\bibitem{BaFe2As2}X. F. Wang, T. Wu, G. Wu, H. Chen, Y. L. Xie, J. J. Ying, Y. J. Yan, R. H. Liu, and X. H. Chen, Phys. Rev. Lett. \textbf{102}, 117005 (2009).


\end{thebibliography}
\end{document}